\documentclass[footinbib,twocolumn,showpacs,amsmath,amstex,amssymb,mathfonts,superscriptaddress,prb]{revtex4-1}
\usepackage{graphicx}
\usepackage{color}
\usepackage{bm}

\usepackage{graphicx}
\usepackage{color}
\usepackage{bm}
\usepackage{amsmath}
\usepackage{amssymb}
\usepackage{amsthm}
\usepackage{amsfonts}
\usepackage{multirow}
\usepackage{makecell}
\usepackage{float}
\usepackage{comment}
\usepackage{enumitem}
\usepackage{tipa}

\usepackage{bbm}

\begin{document}

\title{Effective Abelian theory from a non-Abelian topological order in $\nu=2/5$ fractional quantum Hall effect}
\author{Bo Yang}
\affiliation{Division of Physics and Applied Physics, Nanyang Technological University, Singapore 637371.}
\affiliation{Institute of High Performance Computing, A*STAR, Singapore, 138632.}
\author{Ying-Hai Wu}
\affiliation{School of Physics and Wuhan National High Magnetic Field Center, Huazhong University of Science and Technology, Wuhan 430074, China.}
\author{Zlatko Papi\'c } 
\affiliation{School of Physics and Astronomy, University of Leeds, Leeds LS2 9JT, UK.}


\date{\today}
\begin{abstract}
Topological phases of matter are distinguished by topological invariants, such as Chern numbers and topological spins, that quantize their response to electromagnetic currents and changes of ambient geometry. Intriguingly, in the $\nu=2/5$ fractional quantum Hall effect, prominent theoretical approaches -- the composite fermion theory and conformal field theory -- have constructed two distinct states, the Jain composite fermion (CF) state and the Gaffnian state,
 for which many of the topological indices coincide and even the microscopic ground state wave functions have high overlap with each other in system sizes accessible to numerics. At the same time, some aspects of these states are expected to be very different, e.g., their elementary excitations should have either Abelian (CF) or non-Abelian (Gaffnian) statistics. In this paper we investigate the close relationship between these two states by considering not only their ground states, but also the low-energy charged excitations. We show that the low-energy physics of  both phases is spanned by the same type of quasielectrons of the neighbouring Laughlin phase. The main difference between the two states arises due to an implicit assumption of short-range interaction in the CF approach, which causes a large splitting of the variational energies of the Gaffnian excitations.  
We thus propose that the Jain phase emerges as an effective Abelian low-energy description of the Gaffnian phase when the Hamiltonian is dominated by two-body interactions of sufficiently short range. 
\end{abstract}

\maketitle 

\section{Introduction}\label{sec:intro}

The fractional quantum Hall (FQH) states are a family of strongly correlated electronic states realized in experiments on two-dimensional (2D) semiconductors in strong magnetic fields~\cite{Tsui-PhysRevLett.48.1559}. Their distinct experimental signature -- the robust quantization of Hall conductance -- is a consequence of the underlying  topological invariant~\cite{TKNN}, which also represents the electronic filling factor $\nu$ of the relevant partially-occupied Landau level~\cite{prangegirvin}.
FQH states are  paradigmatic examples of topological phases of matter~\cite{WenTopo} whose low-energy description is governed by topological quantum field theory~\cite{ZHK,LopezFradkin}. Apart from filling factors, these phases are further characterized by non-trivial ground state degeneracy~\cite{WenPhysRevB.41.9377}, modular matrices~\cite{RevModPhys.80.1083}, protected edge excitations~\cite{WenEdge}, topological spin or ``shift"~\cite{wenzee} (which quantizes the dissipationless Hall viscosity~\cite{Avron, ReadViscosity, HaldaneViscosity}) and chiral central charge, which characterizes their response in the thermal Hall effect~\cite{KaneFisher, ReadGreen, CAPPELLI2002}. 

The focus of this paper is the FQH state at filling factor $\nu=2/5$, which is typically the state with the second largest incompressibility gap after the $\nu=1/3$ Laughlin state and its particle-hole conjugate~\cite{StormerRMP}. In most experiments performed at fairly large magnetic fields,  it is generally believed that the $\nu=2/5$ FQH state is a spin polarized state~\cite{Du1995,Kukushkin1999,Park1998}. The underlying physics of the $\nu=2/5$ state was phenomenologically first explained as a hierarchical state~\cite{Haldane1983,HalperinHierarchy}, i.e., a condensate of quasielectron excitations of the $\nu=1/3$ Laughlin state. An alternative interpretation of the $\nu=2/5$ state comes from the composite fermion (CF) theory~\cite{Jain:1989p294,jainbook}), by which the state can be understood as an integer quantum Hall fluid of composite fermions -- bound states consisting of one electron and two quantized vortices. The $\nu=2/5$ state of electrons can be mapped onto two completely filled CF levels which can be roughly viewed as analogs of Landau levels for electrons and they will be referred to as ``$\Lambda$'' levels below. Both the hierarchical approach and the CF approach produce trial wave functions~\cite{Haldane1983, Jain:1989p294,Greiter1994,HanssonRMP} for the ground state at $\nu=2/5$ which have high overlaps with each other as well as with the exact ground state of Coulomb interaction projected into the lowest LL (LLL). Neutral excitation spectrum at $\nu=2/5$ consists of a gapped collective mode~\cite{Golkar,Balram2017}, while charged excitations are also gapped and obey Abelian braiding statistics~\cite{JainStatistics}.

Interestingly, another trial wave function  under the name ``Gaffnian"~\cite{SimonGaffnian} has been constructed for  the $\nu=2/5$ state. This wave function has many elegant algebraic properties which the CF state lacks: (i) it is a correlator of primary fields in a minimal model $M(3,5)$ in conformal field theory (CFT)~\cite{yellow}; (ii) its first-quantized representation is a Jack polynomial~\cite{jack} with parameters $(k=2,r=3)$ that encode the clustering condition which allows exactly up to two electrons in every five consecutive LL orbitals; (iii) the Gaffnian wave function is exactly annihilated by a certain 3-body local Hamiltonian~\cite{SimonRezayiCooper}. Properties (i)-(iii) are  reminiscent of incompressible FQH states, such as the Read-Rezayi states~\cite{ReadRezayiParafermion}, which includes the Laughlin~\cite{Laughlin-PhysRevLett.50.1395} and Moore-Read~\cite{Moore1991362} states. Similar to the Moore-Read case, the combination of (i) and (iii) predicts that charged excitations of the Gaffnian should have non-Abelian braiding statistics~\cite{SimonGaffnian} (see also Ref.~\onlinecite{Flavin}). However, in contrast to these examples, the Gaffnian CFT is \emph{non-unitary}. This implies, via bulk-boundary correspondence, anomalous behaviour at the edge of the system~\cite{ReadConformalInvariance}, which is clearly at odds with the seemingly well-behaved bulk wave function. This ``paradox'' would be resolved if the Gaffnian parent Hamiltonian had \emph{gapless} energy spectrum~\cite{ReadViscosity}, potentially arising as a critical point between two gapped FQH phases (such as in a bilayer system with tunneling~\cite{Papic332}). Recent numerical tests on the sphere~\cite{JolicoeurGaffnian} are indeed consistent with the Gaffnian neutral gap vanishing with system size, although similar results on the torus geometry have been inconclusive~\cite{KangGaffnian}. Furthermore, it has been analytically established that in the so-called thin torus limit the Gaffnian state is gapped~\cite{PapicSolvable,SeidelGaffnian}.  

These differences between the Jain and Gaffnian states suggest that they are different topological phases of matter. However, it was noted early on~\cite{SimonGaffnian} that the two states have surprisingly large (in excess of 95\%) overlap with each other in finite systems where they can be exactly compared~\cite{RegnaultGaffnian}. Both of them also have high overlap with the ground state of Coulomb interaction projected to the LLL. Furthermore, they they share a dominant low-lying part of the entanglement spectrum~\cite{LiHaldane, RegnaultGaffnian}, which describes virtual excitations across a bipartition of a system in orbital space. Moreover, the so-called root partition, which defines the structure of a FQH state under ``squeezing"~\cite{jack, ThomaleProduct}, is closely related for the Jain and Gaffnian states~\cite{RegnaultGaffnian, MilovanovicGaffnian}. Finally, an intriguing recent result~\cite{EstienneGaffnian} suggests that the topological entanglement entropy $\gamma$~\cite{KitaevPreskill,LevinWen}, measured from the entanglement spectrum of the Gaffnian state obtained using exact matrix product state representation of the wave function~\cite{ZaletelMong,EstienneMPS1,EstienneMPS2}, takes the Abelian value of the Jain state. (We note that similar conclusion was reached for the Haldane-Rezayi spin-singlet state~\cite{HR} at $\nu=1/2$, which is also governed by a non-unitary CFT~\cite{Crepel2019}.) These observations raise the question precisely how the two theories, Jain and Gaffnian, are different from each other, in particular from the point of view of data that can be measured in finite-size calculations.      

In this paper we investigate in detail the connection between Jain and Gaffnian states from a standpoint of the effective Hilbert space structure that describes their low-energy physics. As we explain in detail in Sec.~\ref{sec:ham} below, our motivation is to avoid invoking the notion of a parent Hamiltonian when we compare the two phases, because this would entail discussing the energetics of excitations, which is fundamentally a non-topological property. Instead, our approach will rely on the the recently introduced notion of ``local exclusion conditions" (LEC)~\cite{Yang2018} that we briefly review in Sec.~\ref{sec:lec}. The analysis within the LEC framework allows us to focus on topological indices that can be computed from the wave functions themselves \emph{without} referring to a specific model Hamiltonian. In this particular sense, we argue that the Gaffnian phase and the Jain phase at $\nu=2/5$ are ``topologically equivalent" upon restriction to the low-energy subspace. Our extensive numerical analysis presented in Secs.~\ref{sec:gs} and \ref{sec:qe} shows that the Gaffnian ground state and the Jain $\nu=2/5$ state are made of the same type of Laughlin quasielectrons. There is also a close relationship between the Gaffnian quasihole states, and the quasihole states constructed in the CF picture at $\nu=2/5$. In addition, the low-lying quasielectron excitations of the Gaffnian state can be defined from the Hilbert space algebra alone. We show they qualitatively agree with the CF quasielectron states obtained by adding CFs in the third $\Lambda$ level on top of the Jain state.  

The analysis of the quasiholes and quasielectrons of the Gaffnian phase also allow us to understand the incompressibility and topological properties (e.g., the braiding statistics of charged excitations) of the Gaffnian from a microscopic perspective with realistic interactions. As we discuss in Sec.~\ref{sec:inc}, the results imply that the robustness of a topological phase is not only determined by the ground state gap in the thermodynamic limit, but also by the energy splitting of the quasihole manifold under realistic experimental conditions. The Gaffnian phase is non-Abelian due to the structure of its quasihole manifold, resulting in multiple linearly independent states when the positions of quasiholes are fixed. The Jain phase is an effective low energy description with realistic interactions (e.g., the LLL-projected Coulomb interaction), which captures only the low-lying part of the Gaffnian quasiholes, thus effectively rendering the phase Abelian.

\section{The notion of topological equivalence based on the Hilbert space structure}\label{sec:ham}

A traditional approach in condensed matter physics for distinguishing phases of matter is adiabatic continuity, which relies on the behaviour of the energy spectrum under small variations of the Hamiltonian.  An approximate model Hamiltonian for the (LLL-projected) Jain state is the $V_1$ Haldane pseudopotential~\cite{Haldane1983}, which is the leading component of Coulomb interaction in the LLL. On the other hand, the Gaffnian state has an exact Hamiltonian which annihilates it, $H_{\rm Gaff}|\psi_{\rm Gaff}\rangle = 0$, and this Hamiltonian is explicitly given by 
\begin{equation}\label{eq:hgaff}
H_{\rm Gaff} = c_3 H_{\rm 3b}^{M=3} +c_5 H_{\rm 3b}^{M=5},
\end{equation}
where $H_{\rm 3b}^M$ is a projection operator~\cite{SimonRezayiCooper, CHLeePapicThomale} onto states of 3 electrons with relative angular momentum $M$, and $c_3, c_5>0$ are arbitrary positive constants. Numerics suggest~\cite{JolicoeurGaffnian}  $H_{\rm Gaff}$ is gapless only in the thermodynamic limit, while there is a gap in any finite size system. Increasing the system size does not lead to level crossing, and the Gaffnian model wave function is always the zero energy ground state for the positive semidefinite Hamiltonian in Eq.~(\ref{eq:hgaff}). On the other hand, adding any finite amount of two-body $V_1$ pseudopotential to $H_{\rm Gaff}$ appears to increase the ground state gap~\cite{JolicoeurGaffnian}, and the Gaffnian and Jain states can be smoothly deformed into each other in any finite system~\cite{TokeGaffnian}. One may attempt to circumvent the difficult computation of the Hamiltonian gap by studying the decay of correlations in the ground state wave function. Extensive work on one-dimensional systems has shown that when such systems are gapped, they generically exhibit exponentially decaying correlations in the ground state~\cite{Hastings2006}, while gapless systems have power-law decaying correlations~\cite{LSM}. Although these results are expected to apply to two-dimensional systems, they have not been rigorously proven. In addition, a much larger difficulty of simulating 2D systems typically prevents to reliably distinguish power law from exponential decay of correlations. 

In this paper we put aside the adiabatic continuity approach which is based on a difficult problem of bounding the gap of many-body Hamiltonians, and instead we focus on the topological indices of the $\nu=2/5$ state. More specifically, whether or not the Gaffnian state and the Jain $\nu=2/5$ state are topologically equivalent can be determined, at least in principle, in a precise manner: we look at all topological indices that can be computed from the wave functions alone. These topological indices are all integers when measured in appropriate units. If there exists a topological index that takes different values when computed from the two wave functions, then we can unambiguously claim that the Gaffnian state is topologically distinct from the Jain state. Two states with different topological indices also cannot be connected adiabatically by smooth tuning of Hamiltonians in principle.

 At this point, we would like to note a technical difference between two types of topological indices: those that can be computed \emph{exactly} in a finite system vs. those that must be obtained via extrapolation to the thermodynamic  limit. Some examples of the former include the filling factor $\nu=2/5$ and orbital shift $\mathcal{S}=-3$, which are identical for the Jain and Gaffnian states. The same shifts  imply, in particular, that the Jain and Gaffnian states should have the same Hall viscosity~\cite{ReadViscosity,RRViscosity,FremlingViscosity}. An example of a topological index that can be obtained via extrapolation to the thermodynamic limit is the topological entanglement entropy $\gamma=\ln \mathcal{D}$, which is given by the logarithm of the so-called total quantum dimension $\mathcal{D}$ of the CFT~\cite{yellow}. $\gamma$ can be extracted from the scaling of the bipartite entanglement entropy, $S = \alpha L - \gamma + \mathcal{O}(e^{-L})$, where $L$ is the length of the bipartition and $\alpha$ is a non-universal constant prefactor. The entanglement entropy of the Gaffnian has been computed from an exact matrix-product state representation of its wave function on an infinite cylinder~\cite{EstienneGaffnian}, where  ``finite size" effects are conveniently replaced by a ``finite entanglement" truncation of the MPS. Careful scaling of $\gamma$ as a function of entanglement truncation revealed a value that surprisingly matches that of the \emph{Abelian} theory which describes the Jain CF state. This result points to the need for closer scrutiny of the relationship between Jain and Gaffnian states.  Apart from $\gamma$, there is another topological index that in principle distinguishes the Jain and Gaffnian states: the chiral central charge, $c_{-}$. This quantity can also be obtained from the entanglement spectrum and we will discuss it in more detail in Sec.~\ref{sec:c}. 
 
In addition to topological quantum numbers, below we will also examine the complete algebra of the Hilbert space that describes the low-energy physics of Jain and Gaffnian states. By this we mean the space spanned by the ground states and low-lying excitations of the two states, i.e., single and multiple quasihole/quasielectron excitations. We emphasize that these  can be obtained without specifying a Hamiltonian. For example, within the CF theory, these states are generated by explicitly promoting CFs into higher $\Lambda$ levels, and by subsequently  projecting into the LLL; these procedures can be implemented without specifying the Hamiltonian (although, as we will explain below, in practice they are often performed with an implicit assumption of the target Hamiltonian whose properties the states are supposed to model). On the other hand, in the Gaffnian case, we can generate the relevant low-energy Hilbert space from the clustering conditions using an approach of ``local exclusion conditions" (LECs), which was recently introduced in Ref.~\onlinecite{Yang2018} and which we briefly review in Sec.~\ref{sec:lec}. These conditions are similar to the notion of clustering that is used to define the root configurations of Jack polynomials~\cite{jack}, but they can also be applied to many FQH states without a Jack description. Thus, the LEC  framework  allows us to examine the properties of excitations more generally, and to avoid talking about the energetics of such excitations, which is a non-topological property. Finally, we note that ground state degeneracy on a torus is also different for the Jain and Gaffnian states~\cite{SimonGaffnian}. However, since the ground states on the torus correspond to different types of anyons in a topological phase~\cite{RevModPhys.80.1083}, our analysis of charged excitations within the LEC framework will implicitly shed light on this aspect of the two states.

\section{Brief overview of Local Exclusion Conditions}\label{sec:lec}

In this work we employ the recently introduced characterization of FQH states via Local Exclusion Conditions (LECs)~\cite{Yang2018}. Each LEC is defined by a triplet of integers
\begin{equation}\label{eq:lec}
\hat n=\{n,n_e,n_h\}.
\end{equation}
This dictates that for a physical measurement on a circular droplet containing $n$ magnetic fluxes anywhere in the quantum Hall fluid, no more than $n_e$ electrons and $n_h$ holes can be detected. This is formally a classical constraint on the reduced density matrix of the circular droplet within the bulk of the quantum Hall fluid. Imposing the LEC in Eq.~(\ref{eq:lec}) yields the wave functions of the ground states and elementary excitations for a large class of known and new FQH ground states\cite{yangqe}. 

For finite systems, the LEC approach is most conveniently implemented on a spherical geometry, where imposing the constraint on the physical measurement anywhere in the quantum Hall fluid translates into demanding that states are also the highest weight eigenvectors of the total angular momentum $\mathbf{L}^2$ operator\cite{Yang2018,yangqe} in the Hilbert space truncated according to Eq.~(\ref{eq:lec}). When we put the fractional quantum Hall states on the sphere with a magnetic monopole of strength $2S$ at the center, the many-body states can be expanded into second quantised bases that are Slater determinants of $N_e$ number of electrons, and $N_o$ number of orbitals ($N_o=2S+1+2N$, where $N$ is the Landau level index)\cite{Haldane1983,jack}. The truncation can be implemented by looking at $n$ orbitals centered at the north pole. These $n$ orbitals naturally form the droplet at the north pole with the physical area of $\sim 2\pi nl_B^2$. Thus the imposition of an LEC in the form of Eq.(\ref{eq:lec}) dictates that no more than $n_e$ electrons or $n_h$ number of holes can be contained in that droplet. 

Taking the Hilbert space of $N_e=6,N_o=16$ and the total z-component angular momentum $L_z=0$ as an example, all basis states are squeezed from $1110000000000111$, and a few bases are listed as follows:
\begin{eqnarray}
&&1110000000000111\label{a1}\\
&&1101000000001011\label{a2}\\
&&1011000000001101\\
&&\qquad\qquad\vdots\nonumber\\
&&0000011111100000
\end{eqnarray}
The string of numbers in the occupation bases from the left to right correspond to orbitals on the sphere from the north pole to the south pole, where $1/0$ indicates the corresponding orbital is occupied/unoccupied. If we take $\hat n=\{2,1,2\}$, which implies the two orbitals at the north pole cannot contain more than one electron, the truncation of the Hilbert space is implemented by looking at the two leftmost orbitals while scanning through all bases in the Hilbert space. If the two leftmost orbitals are both occupied (e.g. (\ref{a1}),(\ref{a2})), such bases will be removed. In this particular case, there is only one highest weight state (by diagonalising the $\mathbf{L}^2$ operator) in the truncated Hilbert space. More generally, only when $N_o=3N_e-2$, we obtain a unique highest weight state in the Hilbert space truncated by $\hat n=\{2,1,2\}$, which turns out to be the Laughlin state. When $N_o<3N_e-2$, no highest weight states exist in any such truncated Hilbert space. This is also how we determine unambiguously the topological indices (e.g the filling factor $\nu=1/3$, the topological shift $\mathcal S=-2$) for the Laughlin FQH phase.

In simple cases, the LEC approach produces FQH states and their excitations that are identical to those from the Jack polynomials approach. However, at general filling factors the LEC construction also admits states which do not coincide with any known Jacks. In this work we focus on the Hilbert space at filling factor $\nu=2/5$, with the number of orbitals set to $N_o = 5N_e/2-3$, where $N_e$ is the number of electrons and $\mathcal{S}=-3$ is the orbital shift of the Jain and Gaffnian states. Explicit numerical calculations show that the algebraic structure of the Hilbert space that results from imposing the specific LECs, 
\begin{eqnarray}\label{eq:cg}
\hat c_G =\{2,1,2\}\lor\{5,2,5\},
\end{eqnarray}
uniquely determines the ground state and the quasihole manifold of the Gaffnian phase. Here the symbol $\lor$ implies the FQH fluid satisfies $\{2,1,2\}$ (any circular droplet containing \emph{two} fluxes does not contain more than \emph{one} electron) \emph{or} $\{5,2,5\}$ (any circular droplet containing \emph{five} fluxes does not contain more than \emph{two} electrons). The same LECs also defines the $\left(5,2\right)\text{Type}$ quasielectrons of the Laughlin phase at $\nu=1/3$. The low-lying quasielectron excitations of the Gaffnian phase, on the other hand, can be defined by 
\begin{eqnarray}\label{eq:ch}
\hat c_h=\{2,1,2\}\lor\{6,3,6\}.
\end{eqnarray}
 We next present detailed comparisons of the ground state and elementary excitations of the Gaffnian phase with the Abelian Jain $\nu=2/5$ phase in the CF picture. 

\section{ Ground states and quasiholes at $\nu=2/5$}\label{sec:gs}

As mentioned in Sec.~\ref{sec:intro}, the relationship between the Gaffnian state and the Jain  state at $\nu=2/5$ is quite subtle. The Gaffnian has the root configuration~\cite{jack} 
\begin{eqnarray}\label{eq:gaffroot}
1100011000\cdots 1100011,
\end{eqnarray}
where $\cdots$ represent repeated patterns of $11000$. In contrast, the Jain state has the root configuration~\cite{RegnaultGaffnian}
\begin{eqnarray}\label{eq:jainroot}
110010010\cdots 010010011
\end{eqnarray}
where $\cdots$ represent repeated patterns of $10010$, which is ``unsqueezed" from that of the Gaffnian~\cite{RegnaultGaffnian}. In both cases, the states only consist of basis vectors squeezed from their respective root configuration and they have very high overlap  in  finite systems. One important difference is that the root configuration Eq.~(\ref{eq:jainroot}) does not uniquely specify the Jain state~\cite{RegnaultGaffnian}.

\begin{figure}[htb]
\includegraphics[width=\linewidth]{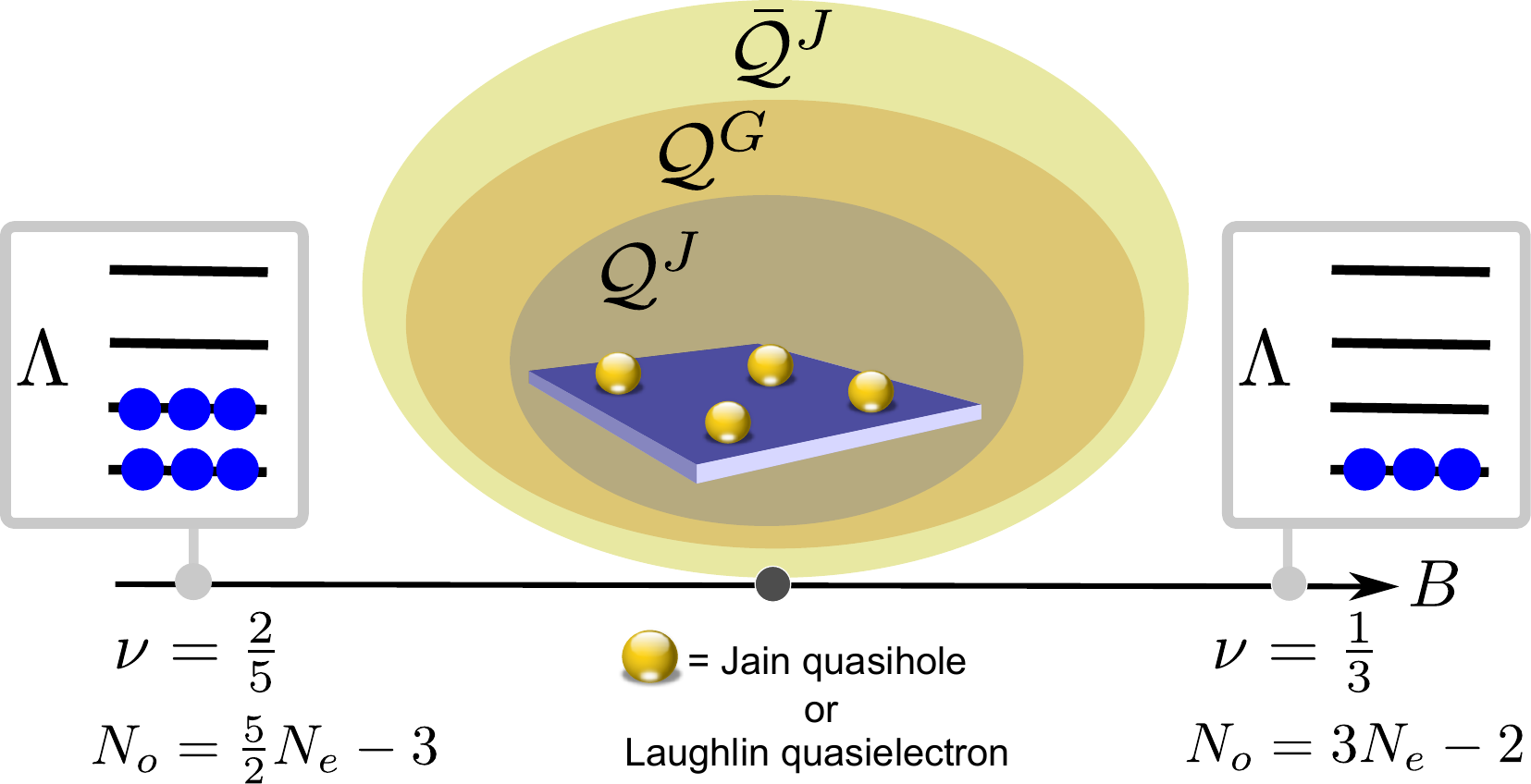}
\caption{An illustration of possible quasihole subspaces of the $\nu=2/5$ state when $ \frac{5}{2}N_e - 3< N_o < 3N_e -2$. The subspace of undressed Jain quasiholes (or Laughlin quasielectrons) $\mathcal{Q}^J$ is contained within the subspace of Gaffnian quasiholes, $\mathcal{Q}^G$. Intriguingly, the Gaffnian subspace itself is contained within the generalized Jain quasihole subspace, $\bar{\mathcal{Q}}^J$.}
\label{fig1}
\end{figure}

Next, we present additional evidence that the two states could be physically equivalent (in the sense of our Sec.~\ref{sec:ham}) by looking at the quasihole states and their connections to the quasielectron states arising from the Laughlin state at $\nu=1/3$. The crucial insight discovered in Ref.~\onlinecite{yangqe} is that the CF quasielectrons in the second $\Lambda$ level is qualitatively equivalent to the $\left(5,2\right)$ Type Laughlin quasielectrons obtained from LEC constructions.~\footnote{Note they are not equivalent to the original Laughlin quasiparticles defined in Ref.~\onlinecite{Laughlin-PhysRevLett.50.1395}.} We analyze this in more detail by studying different Hilbert spaces indexed by $(N_o, N_e)$, in which various possible states live --  see the summary in Fig.~\ref{fig1}.

From the CF perspective, the Laughlin ground state at $\nu=1/3$ occurs at $N_o=3N_e-2$, with the fully occupied lowest $\Lambda$ level. For $N_o<3N_e-2$, the Laughlin quasielectrons are constructed by keeping the lowest $\Lambda$ level fully occupied, and adding CFs to the first $\Lambda$ level. This is a particular type of quasielectron~\cite{Yang14}  related to the familiar magnetoroton or Girvin-MacDonald-Platzman~\cite{GMP86} neutral mode, in which each neutral excitation consists of one quasihole and one quasielectron of this type. The quasielectrons thus constructed are \emph{undressed} quasielectrons. Dressed quasielectrons can be constructed by exciting CFs in the lowest $\Lambda$ level to the second $\Lambda$ level, so that the quasielectrons are dressed with neutral excitations consisting of quasielectron-quasihole pairs\cite{ParkJain,Majumder2009}. The maximum number of CFs to be added to the second $\Lambda$ level is also constrained by $N_o\ge 5N_e/2-3$, beyond which the second CF level is completely filled, and additional CFs can only be added to the third $\Lambda$ level.

The Hilbert space of $N_o= 5N_e/2-3$ corresponds to both the lowest and first $\Lambda$ level completely filled, which is mapped to the Jain $\nu=2/5$ ground state. It is important to note that Hilbert spaces with $3N_e-2>N_o> 5N_e/2-3$ correspond to both the Laughlin quasielectrons and the Jain $\nu=2/5$ quasiholes. In particular, every \emph{undressed} Laughlin quasielectron state (which may contain multiple quasielectrons) is equivalent to a Jain $\nu=2/5$ quasihole state (which may contain multiple quasiholes) in the CF picture. This is because a Jain $\nu=2/5$ quasihole state only involves the removal of CFs in the second $\Lambda$ level~\cite{TokeGaffnian}. We denote the Hilbert space of all Jain $\nu=2/5$ quasielectrons as $\mathcal Q^J$. We can also define the family of ``generalized Jain $\nu=2/5$ quasiholes", which allows the removal of CFs from either the lowest or the first $\Lambda$ level of the Jain $\nu=2/5$ ground state. The generalized Jain $\nu=2/5$ quasiholes are equivalent to Laughlin quasielectron states, both dressed and undressed by neutral excitations, and we denote the spanned Hilbert space as $\bar{\mathcal Q}^J$. We obviously have $\mathcal Q^J\subset\bar{\mathcal Q}^J$, and the generalized Jain $\nu=2/5$ quasiholes are microscopically well-defined objects. 

We now analyze the Hilbert spaces in Fig.~\ref{fig1} from the LEC perspective. At $N_o=3N_e-2$, the unique highest weight state after imposing $\hat c_L=\{2,1,2\}$ is the translationally invariant Laughlin state, which exactly agrees with the CF construction. At $N_o=5N_e/2-3$, the unique translationally invariant state can be obtained by imposing the constraint in Eq.~(\ref{eq:cg}), and the resulting state is the exact Gaffnian state. The important observation here is that for $3N_e-2>N_o> 5N_e/2-3$, the imposition of $\hat c_G$ defines a manifold of highest weight states, which can be identified as the $\left(5,2\right)\text{Type}$ Laughlin quasielectron manifold~\cite{yangqe}. Each state consists of local charge excitations from the Laughlin $\nu=1/3$ ground state. This quasielectron manifold is well-defined from the algebraic structure of the Hilbert space, independent of the microscopic Hamiltonian, and we denote it as $\mathcal Q^G$. Every state in this manifold contains a number of $\left(5,2\right)\text{Type}$ Laughlin quasielectrons and quasiholes. The familiar magnetoroton mode also belongs to this manifold~\cite{Yang14,Majumder2009}. On the other hand, the creation energy of the quasielectrons and quasiholes, as well as the interaction between them, are non-universal and depend on the details of the microscopic Hamiltonian. In particular, the Hamiltonian must be specified to  obtain more information from the states, including the number of quasielectrons and/or the number of quasiholes that effectively remain in the low-energy theory\cite{yangqe}.
\begin{table}[h!]
\centering
\begin{tabular}{ |c|c|c|c|c|} 
 \hline
 $N_e$ & $N_\phi$ & Quasielectrons No. & Overlap & $L$ sector \\ 
 \hline
  $6$ & $15$ & 1 & 0.993191247&3 \\ 
 \hline
   $7$ & $18$ & 1 & 0.992358615&3.5 \\ 
 \hline
  $8$ & $21$ & 1 & 0.991746992&4 \\ 
 \hline
  $9$ & $24$ & 1 & 0.991214478&4.5 \\ 
 \hline
   $6$ & $14$ & 2 & 0.985751939&0 \\ 
 \hline
  $6$ & $14$ & 2& 0.984173168&2\\ 
 \hline
  $6$ & $14$ & 2 & 0.984173168&4 \\ 
 \hline
  $7$ & $17$ & 2 & 0.988597494&1 \\ 
 \hline
  $7$ & $17$ & 2 & 0.983286491&3 \\ 
 \hline
  $7$ & $17$ & 2 & 0.943082997&5 \\ 
 \hline
  $8$ & $20$ & 2 & 0.985628951&0 \\ 
 \hline
  $8$ & $20$ & 2 & 0.986852018&2 \\ 
 \hline
  $8$ & $20$ & 2 & 0.972283336&4 \\ 
 \hline
  $8$ & $20$ & 2 & 0.949012693&6 \\ 
 \hline
  $9$ & $23$ & 2 & 0.986292384&1 \\ 
 \hline
  $9$ & $23$ & 2 & 0.984033466&3 \\ 
 \hline
  $9$ & $23$ & 2 & 0.973425667&5 \\ 
 \hline
  $9$ & $23$ & 2 & 0.953128136&7 \\ 
   \hline
  $10$ & $26$ & 2 & 0.9881478295&0 \\ 
 \hline
   $10$ & $26$ & 2 & 0.9833100134&2 \\ 
 \hline
    $10$ & $26$ & 2 & 0.9827543856&4 \\ 
 \hline
     $10$ & $26$ & 2 & 0.9722622828&6 \\ 
 \hline
      $10$ & $26$ & 2 & 0.9544481270&8 \\ 
 \hline
       $6$ & $13$ & 3 & 0.8743444175&1 \\ 
 \hline
        $6$ & $13$ & 3 & 0.8105834258&3 \\ 
 \hline
         $7$ & $16$ & 3 & 0.9342778754&1.5 \\ 
 \hline
          $7$ & $16$ & 3 & 0.9216743063&2.5 \\ 
 \hline
           $7$ & $16$ & 3 & 0.8775111803&4.5 \\ 
 \hline
      $8$ & $19$ & 3 & 0.9099956938&0 \\ 
 \hline
       $8$ & $19$ & 3 & 0.9319712444&2 \\ 
 \hline
        $8$ & $19$ & 3 & 0.9503537167&3 \\ 
 \hline
         $8$ & $19$ & 3 & 0.9215094248&4 \\ 
 \hline
          $8$ & $19$ & 3 & 0.8952653937&6 \\ 
 \hline
   $9$ & $22$ & 3 & 0.960168064&1.5 \\ 
 \hline
   $9$ & $22$ & 3 & 0.949196167&2.5 \\ 
 \hline
   $9$ & $22$ & 3 & 0.936635226&3.5 \\ 
 \hline
   $9$ & $22$ & 3 & 0.934275046&4.5 \\ 
 \hline
   $9$ & $22$ & 3 & 0.930653326&5.5 \\ 
 \hline
   $9$ & $22$ & 3 & 0.910263617&7.5 \\ 
 \hline
    $10$ & $24$ & 4 & 0.8928222791&0 \\ 
 \hline
     $10$ & $24$ & 4 & 0.9114580216&5 \\ 
 \hline
      $10$ & $24$ & 4 & 0.8842522966&6 \\ 
 \hline
       $10$ & $24$ & 4 & 0.8719329406&7 \\ 
 \hline
\end{tabular}
\caption{Wave function overlap of the $\left(5,2\right)\text{Type}$ Laughlin quasielectrons with the CF quasielectrons in the second $\Lambda$ level (first $\Lambda$ level completely filled). The quasielectron number is the number of $\left(5,2\right)\text{Type}$ Laughlin quasielectrons or the CF quasielectrons }
\label{t1}
\end{table}

The single $\left(5,2\right)\text{Type}$ Laughlin quasielectron state from the LEC construction can be found in the Hilbert space of $N_o=3N_e-3$. We can thus denote $\mathcal Q_{\text{1qe}}^{\left(5,2\right)}$ as the single quasielectron manifold, equivalently the highest weight subspace after imposing $\hat c_G$ on the full Hilbert space. In the Laughlin phase, we can assume the creation energy of the quasiholes are small, while that of the $\left(5,2\right)\text{Type}$ Laughlin quasielectrons are finite, implying incompressibility and representing the charge gap necessary for the phase to be robust. If we diagonalize a short range Hamiltonian (e.g., containing only $V_1$ pseudopotential) within $\mathcal Q_{\text{1qe}}^{\left(5,2\right)}$, the lowest energy state should contain a single undressed quasielectron, while the excited states contain one quasielectron dressed by neutral excitation(s). 

In Table~\ref{t1} we show that a single $\left(5,2\right)\text{Type}$ Laughlin quasielectron has the same quantum number and very high overlap with the single quasielectron state from the CF construction. We thus claim the $\left(5,2\right)\text{Type}$ Laughlin quasielectron is qualitatively equivalent to the Laughlin quasielectron in the CF picture. It is thus reasonable to assume that all states containing multiple undressed $\left(5,2\right)\text{Type}$ Laughlin quasielectrons are equivalent to the undressed Laughlin quasielectron states in the CF picture (see Table~\ref{t1} for a few concrete examples). This implies the Hilbert space spanned by all states containing only undressed $\left(5,2\right)\text{Type}$ quasielectron is also given by $\mathcal Q^J$. In particular, the Jain $\nu=2/5$ ground state is made entirely of undressed Laughlin quasielectrons, and the Gaffnian state is made entirely of undressed $\left(5,2\right)\text{Type}$ quasielectrons. The fact that the Jain $\nu=2/5$ and Gaffnian states are built from the same type quasielectrons of the Laughlin phase is strong evidence that the two states capture the same low-energy physics spanned by states within $\mathcal{Q}^J$.

\begin{table}[h!]
\centering
\begin{tabular}{ |c|c|c|c|c|} 
 \hline
 $N_e$ & $N_\phi$ & Quasielectrons No. & Overlap & $L$ sector \\ 
 \hline
  $6$ & $15$ & 1 & 0.973487198&4 \\ 
 \hline
   $7$ & $18$ & 1 & 0.968581199&4.5 \\ 
 \hline
  $8$ & $21$ & 1 & 0.965443776&5 \\ 
 \hline
  $9$ & $24$ & 1 & 0.963056386&5.5 \\ 
 \hline
\end{tabular}
\caption{Wave function overlap of the $\left(6,3\right)\text{Type}$ Laughlin quasielectrons with the CF quasielectrons in the third $\Lambda$ level (first $\Lambda$ level completely filled).}
\label{t2}
\end{table}
We now turn our focus to the $\nu=2/5$ quasihole manifold. In the CF picture, both the generalized Jain quasiholes $\bar{\mathcal Q}^J$ and the conventional quasiholes $\mathcal Q^J$ can be reinterpreted as Laughlin quasielectrons (dressed and undressed). Similarly, the $\left(5,2\right)\text{Type}$ Laughlin quasielectron manifold $\mathcal Q^G$ comes from the highest weight states with $\hat c_G$, so $\mathcal Q^G$ is the quasihole manifold of the Gaffnian state, i.e., all states in $\mathcal Q^G$ have zero energy with the Gaffnian model Hamiltonian in Eq.~(\ref{eq:hgaff}). It is easy to see in the lowest LL with Coulomb interaction, or with any interaction dominated by $V_1$ pseudopotential, that the finite energy cost of creating a single $\left(5,2\right)\text{Type}$ quasielectron in the thermodynamic limit dictates that different Gaffnian quasiholes have different variational energies: quasiholes corresponding to dressed quasielectron states have higher energies than those corresponding to the undressed ones, as each neutral excitation also costs a finite amount of energy.

We have therefore defined three sub-Hilbert spaces of the quasihole excitations: the conventional Jain $\nu=2/5$ quasihole manifold $\mathcal Q^J$, obtained by the removal of composite fermions from only the second $\Lambda$ level (fully occupied for the $\nu=2/5$ ground state); the generalised Jain $\nu=2/5$ quasihole manifold $\bar{\mathcal Q}^J$, obtained by the removal of composite fermions from either the second $\Lambda$ level or the first $\Lambda$ level or both; the Gaffnian quasihole manifold $\mathcal Q^G$, obtained from the LEC approach with $\hat c_G$. All three quasihole sub-Hilbert spaces are from $N_o>2N_e/5-3$. From the extensive numerical analysis, it turns out the various quasihole manifolds have the following relationship depicted in Fig.~\ref{fig1}:
\begin{eqnarray}
\mathcal Q^J \subset \mathcal Q^G \subset \bar{\mathcal Q}^J.
\end{eqnarray}
The Gaffnian quasiholes are non-Abelian, and $\mathcal Q^J$ is a subspace of $\mathcal Q^G$. It is intriguing that $\mathcal Q^G$ is a subspace of $\bar{\mathcal Q}^J$, implying that the Jain $\nu=2/5$ phase could be non-Abelian (with higher intrinsic degeneracy for multiple quasihole states than the Gaffnian quasiholes). It is thus important to note that the Abelian nature of the Jain $\nu=2/5$ phase is crucially dependent on only allowing CFs in the second $\Lambda$ level to be removed to create quasiholes. This is only justified in the presence of short range interactions, whereby removing CFs in the lowest $\Lambda$ level gives states with significantly higher variational energies~\cite{TokeGaffnian}. If there exist microscopic Hamiltonians that maintain incompressibility, while at the same time do not energetically penalise the removal of CFs in the lowest $\Lambda$ level (i.e., the lowest and the first $\Lambda$ levels become quasi-degenerate in the thermodynamic limit), then $\bar{\mathcal Q}^J$ should be the proper quasihole manifold for the Jain state. 

Another possibility is that the additional states in $\bar{\mathcal Q}^J$, as compared to those in $\mathcal Q^G$ are due to the uncontrolled process of LL projection, or strong interactions between the CFs that are not fully accounted for. This can be particularly illustrated by considering the case of the single quasihole state of the Jain $\nu=2/5$ phase. In the CF theory, the state can be intuitively constructed with $N$ CFs in the lowest $\Lambda$ level, and $N+1$ CFs in the first $\Lambda$ level. This corresponds to the removal of a single CF from the $\nu=2/5$ ground state. However, there appears to be no \emph{single quasihole state} for the Gaffnian phase (contrary to the claim in Ref.~ \cite{TokeGaffnian}). For example, with $N_e=7, N_o=15$, there is no zero energy eigenstate of the Gaffnian model Hamiltonian. There is also no highest weight state when the Hilbert space is truncated by $\hat c_\alpha=\{2,1,2\}\lor\{5,2,5\}$. The single quasihole for the Jain state occurs in the $L=2$ sector. The corresponding root configuration for the Gaffnian state is given as follows:
\begin{eqnarray}\label{7root}
110001100011\d{0}0\textsubring{1}\textsubring{}\quad\left(5,2\right)\text{Type}\quad L=2
\end{eqnarray}
The black dot beneath the numbers gives the location of a quasiparticle of charge $-e/5$ (when five consecutive orbitals contain more than two electrons), and the circles give the location of the quasiholes of charge $e/5$ (when five consecutive orbitals contain less than two electrons). Thus, in this Hilbert space there is no state with a single quasihole, and the lowest energy state contains a single quasihole dressed by a neutral excitation (of a quasihole-quasiparticle pair). 
Naively, this looks like an example where the counting between the $\left(5,2\right)\text{Type}$ Laughlin quasielectrons and the CF quasielectron fails to match. For the Laughlin $1/3$ ground state with $N_e$ electrons, we cannot add $N_e+1$ $\left(5,2\right)\text{Type}$ Laughlin quasielectrons on top of it, since such a state does not exist from the LEC construction. However, we can add $N_e+1$ CF quasielectrons to the second $\Lambda$ level, as there exists one such state in $L=\left(N_e+1\right)/2$ sector. 

This paradox can be resolved by the fact that the lowest energy state in that sector can be constructed within the LEC formalism by using $\hat c_h=\{2,1,2\}\lor\{6,3,6\}$ (see also Sec.~\ref{sec:qe} below). It thus consists of a Gaffnian quasihole dressed by a neutral excitation consisting of a Gaffnian quasihole and a $\left(6,3\right)\text{Type}$ Gaffnian quasielectron. This should correspond in the CF construction to the configuration with $N_e$ CFs in the lowest $\Lambda$ level, $N_e$ CFs in the second $\Lambda$ level, and one CF in the \emph{third} $\Lambda$ level (also see Table.~\ref{t2}). Indeed, for $N_e=3$ (corresponding to the case given by Eq.(\ref{7root})), the lowest energy state with such configuration in the $L=2$ sector has overlap of $0.997$ with the single quasihole Jain $\nu=2/5$ state (which is given by 3 CFs in the lowest $\Lambda$ level, and 4 CFs in the second $\Lambda$ level). It turns out in the CF construction, the single quasihole state (with all $N_e+1$ CFs in the second $\Lambda$ level) is physically identical to the dressed quasihole state (with $N_e$ CFs in the second CF level, one CF in the third CF level) in the sector $L=\left(N_e+1\right)/2$, most probably due to the fact that interactions between CFs mixes different CF levels.

Similarly, many states in $\bar{\mathcal Q}^J$ actually contain quasielectrons in the third and higher CF levels, which is supported by numerical calculations. Thus the quasihole manifold for the Gaffnian phase and the Jain $\nu=2/5$ phase could also be equivalent with $\mathcal Q^G=\bar{\mathcal Q}^J$, and the two phases differ not because of the topological distinctions of the ground state or the quasihole manifold, but because of the splitting of the quasihole manifold with realistic interactions. We discuss this point in Sec.~\ref{sec:inc} below. 

\section{ Quasielectrons of the $\nu=2/5$ phase}\label{sec:qe}

It is also important to study the types of quasielectron states the topological phase at $\nu=2/5$ can support, and to see if there are qualitative differences between the Gaffnian and Jain descriptions. The construction of $\nu=2/5$ quasielectron states is straightforward in the CF picture: one can just add CFs to the third $\Lambda$ level, given the lowest two $\Lambda$ levels are completely occupied. From the perspective of conformal field theory, the construction of quasielectrons is rather involved~\cite{HanssonRMP} and there are no available results for the Gaffnian case. However, the LEC formalism allows a very natural construction. While the condition in Eq.~(\ref{eq:cg}) uniquely defines the Gaffnian ground state, Gaffnian quasihole manifold and all of its topological properties, a more relaxed LEC of Eq.~(\ref{eq:ch})  leads to the construction of low-lying charged excitations on top of the Gaffnian ground state. This is similar in spirit to how low-lying charged excitations on top of Laughlin $\nu=1/3$ ground state are built. 

We term such quasielectrons constructed by imposing $\hat c_h$ on the Hilbert space with $N_o<5N_e/2-3$ as $\left(6,3\right)\text{Type}$ Gaffnian quasielectrons. A state with a single such quasielectron has to contain an odd number of electrons. Following the approach of~\cite{Yang12,Yang14}, the root configuration for a single $\left(6,3\right)\text{Type}$ Gaffnian quasielectron is given as follows:
\begin{eqnarray}\label{gq}
1110000110001100011000\cdots, \qquad L=\frac{1}{4}\left(N_e+3\right).
\end{eqnarray}
Let $|\psi_{\text{qe}}\rangle$ be the quasielectron state containing only the basis squeezed from Eq.~(\ref{gq}). $|\psi_{\text{qe}}\rangle$ can be uniquely determined by the following two constraints:
\begin{eqnarray}\label{gqc}
L^+|\psi_{\text{qe}}\rangle=0,\qquad \hat H_{\rm Gaff} c_1c_2c_3|\psi_{\text{qe}}\rangle=0,
\end{eqnarray}
where $\hat H_{\rm Gaff}$ was given in Eq.~(\ref{eq:hgaff}) and $c_i$ is the annihilation operator for the electron in the $i^{\text{th}}$ orbital (orbital 1 denotes one of the poles of the sphere). It is difficult to generalise this root configuration approach to cases with multiple quasielectrons. The LEC construction naturally gives the $\left(6,3\right)\text{Type}$ Gaffnian quasielectron manifold including single and multiple quasielectron states, though it requires a Hamiltonian to resolve states with different number of quasielectrons. Let us define $\bar{\mathcal Q}^{\left(6,3\right)}_{\text{1qe}}$ to be the subspace of a single $\left(6,3\right)\text{Type}$ Gaffnian quasielectron (which may or may not be dressed), obtained from the Hilbert space with odd $N_e$ and $N_o=5\left(N_e-1\right)/2-1$. The state with a single undressed $\left(6,3\right)\text{Type}$ Gaffnian quasielectron can again be resolved as the lowest energy state by diagonalising $V_1$ interaction or the Gaffnian model Hamiltonian within $\bar{\mathcal Q}^{\left(6,3\right)}_{\text{1qe}}$. We have checked the overlaps of this state with the state constructed from Eq.(\ref{gq}) and Eq.(\ref{gqc}), as well as the state from the CF construction. The very high overlap in most cases suggests they are all qualitatively equivalent (see Table.~\ref{t3}).
\begin{table}[h!]
\centering
\begin{tabular}{ |c|c|c|c|c|} 
 \hline
 $N_e$ & $N_\phi$ & Quasielectrons No. & Overlap & $L$ sector \\ 
 \hline
  $5$ & $9$ & 1 & 0.992059308&2 \\ 
 \hline
   $7$ & $14$ & 1 & 0.986556772&2.5 \\ 
 \hline
  $9$ & $19$ & 1 & 0.983445077&3 \\ 
 \hline
   $6$ & $11$ & 2 & 0.7752216578 &1 \\ 
 \hline
    $6$ & $11$ & 2 & 0.7881834641  &3 \\ 
 \hline
     $8$ & $16$ & 2 & 0.9890547151  &0 \\ 
 \hline
      $8$ & $16$ & 2 & 0.9218344596  &2 \\ 
 \hline
      $8$ & $16$ & 2 &  0.8983693223  &4 \\ 
 \hline
        $9$ & $18$ & 2 &  0.8530680855  &1.5 \\ 
 \hline
       $9$ & $18$ & 2 &  0.8008992103  &2.5 \\ 
 \hline
        $9$ & $18$ & 2 &   0.6834199602  &4.5 \\ 
 \hline
\end{tabular}
\caption{Wave function overlap of the $\left(6,3\right)\text{Type}$ Gaffnian quasielectrons with the CF quasielectrons in the third $\Lambda$ level (first two $\Lambda$ levels completely filled).}
\label{t3}
\end{table}

\section{Chiral central charge}\label{sec:c}

As noted in the Introduction, the chiral central charge in principle distinguishes the Jain and Gaffnian states. For the Abelian Jain state, the chiral central charge is $c_- = 2$, i.e., it trivially counts the number of chiral edge modes. For the Gaffnian state, in addition to one bosonic mode for the charge sector, there is a neutral sector which is described by the non-unitary $M(3,5)$ minimal model, whose central charge is equal to $-3/5$~\cite{yellow}. Thus, the total central charge of the Gaffnian is $c_- = 1 - 3/5$. Extracting the value of $c_-$ from a wave function is non-trivial but can be accomplished using ``momentum polarization" which was introduced in Ref.~\onlinecite{Tu2013} (see also appendices of Ref.~\onlinecite{ZaletelMongPollmann}). Physically, momentum polarization measures the momentum carried by a subsystem, thus it can be practically evaluated from the knowledge of the orbital entanglement spectrum~\cite{ParkHaldane}. 

Specifically, momentum polarization can be expressed as~\cite{ParkHaldane}
\begin{eqnarray}\label{eq:mompol}
\langle \Delta M_L \rangle =  \frac{\gamma}{24} - h_a + \frac{1}{2}\left( \frac{L}{2\pi\ell_B} \right)^2 \times \frac{-s}{q}.
\end{eqnarray}
On the left hand side, $\Delta M_L$ is the momentum in the left subsystem, measured relative to the root configuration, which can be computed solely by averaging over the entanglement spectrum. The right hand side stipulates that  $\Delta M_L$ scales super-extensively with the length of the bipartition $L$ (in units of the magnetic length, $\ell_B$), with a universal coefficient $-s/q$, where $s$ is the guiding center spin~\cite{HaldaneViscosity} ($s=-3$ for the Gaffnian) and $q$ is the number of orbitals in the unit cell of the root partition ($q=5$ for the Gaffnian). The subleading constant term, $\gamma/24 - h_a$, is also universal, where $h_a$ represents the conformal spin of an anyon of type $a$  in the theory, and $\gamma=c_- - \nu$ is the desired coefficient that depends on the chiral central charge. 

\begin{figure}
\includegraphics[width=\linewidth]{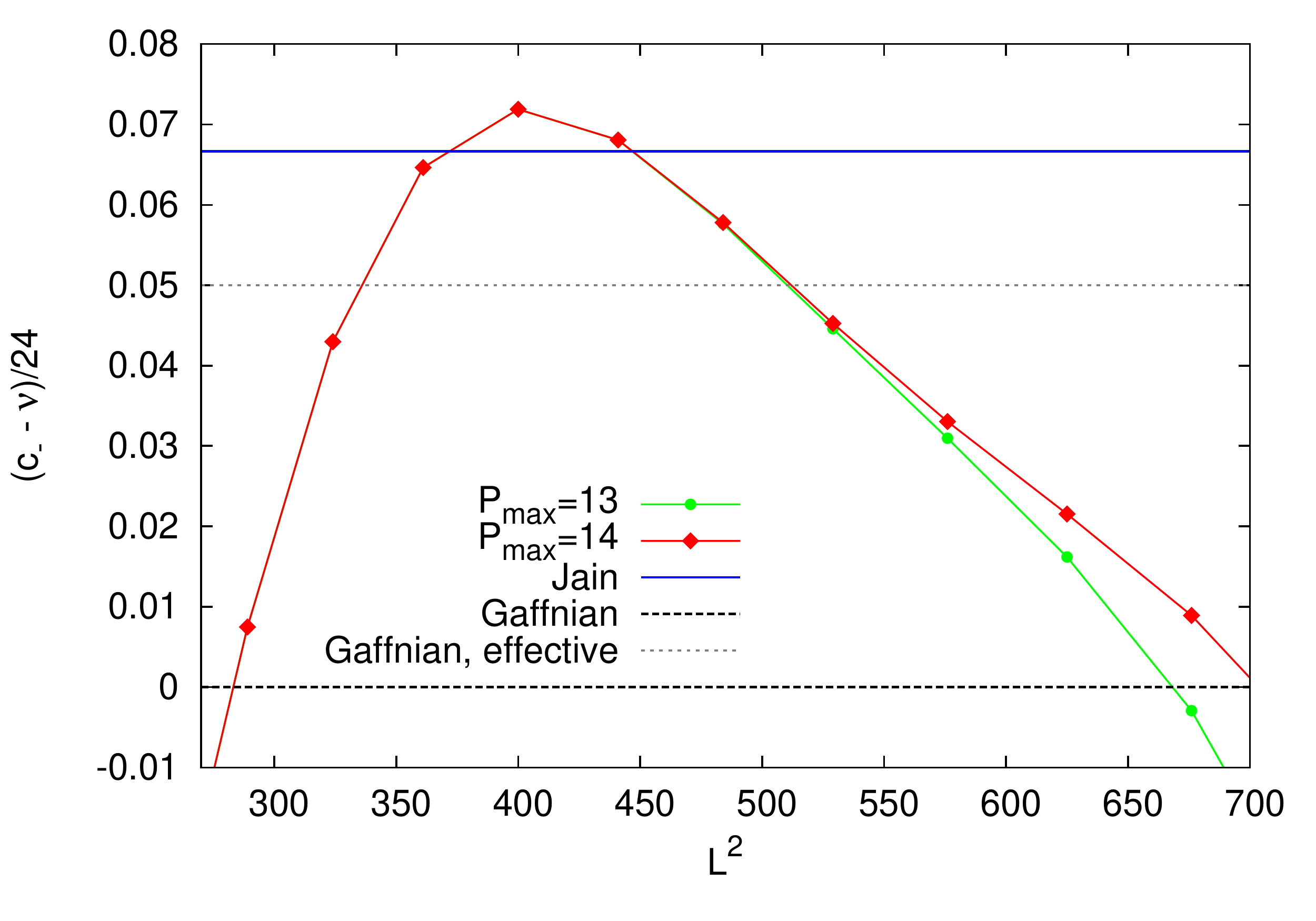}
\caption{Extracting chiral central charge from the exact Gaffnian wave function via momentum polarization in Eq.~(\ref{eq:mompol}). We compute the momentum polarization using the entanglement spectrum of the exact Gaffnian state on the infinite cylinder from Ref.~\onlinecite{EstienneGaffnian}. The entanglement spectrum is calculated for two different levels of truncation, denoted by $P_{\rm max}$. Solid and dashed lines indicate the expected values for the Jain and Gaffnian states, respectively. Dotted line is the ``effective" central charge of the Gaffnian~\cite{JackProperties,ReadConformalInvariance}. }
\label{fig:mompol}
\end{figure}
For the familiar examples of Laughlin and Moore-Read states, the formula in Eq.~(\ref{eq:mompol}) yields the expected values $c_-=1$ and $c_-=3/2$, respectively, in agreement with CFT~\cite{ParkHaldane}. However, a reliable extraction of $c_-$ requires large systems. Following Ref.~\cite{ParkHaldane}, we use Eq.~(\ref{eq:mompol}) in order to extract $c_-$ for the Gaffnian state and the result is shown in Fig.~\ref{fig:mompol}. We have computed momentum polarization from the largest available entanglement spectra of the exact Gaffnian state that was obtained in the MPS representation in Ref.~\onlinecite{EstienneGaffnian}. The MPS computation is controlled by a truncation level $P_{\rm max}$, which roughly corresponds to the number of momentum-resolved sectors of the entanglement spectrum (higher value of $P_{\rm max}$ therefore corresponds to higher accuracy). The FQH system is placed on an infinite cylinder and $L$ denotes the circumference of the cylinder (i.e., the length of the entanglement bipartition). For simplicity, we show the data for the vacuum sector ($h=0$). 

Unfortunately, Fig.~\ref{fig:mompol} does not allow us to draw a definitive conclusion whether the data agrees with the prediction of the Jain (solid line) or Gaffnian (dashed line).  
Comparing with the Moore-Read case in Ref.~\cite{ParkHaldane}, we expect only a limited range of $L$ where the result is fully converged for the values of $P_{max}$ numerics can access. For example, for the Moore-Read case, it is only in the interval $400 \lesssim L^2 \lesssim 550$ that $c_-$ matches the expected value (although it is seen that this interval expands with the increase of $P_{max}$).  Referring back to Fig.~\ref{fig:mompol}, we see that in the similar range of $L$, the value of $c_{-}$ is closer to the prediction of Jain theory, which is consistent with the finding of Ref.~\cite{RegnaultGaffnian}. However, the value of the ``effective" central charge of the Gaffnian~\cite{ReadConformalInvariance}, $c_{\rm eff}=1+3/5$, which can be extracted from the Jack wave function~\cite{JackProperties}, is closer to the expected value for the Jain state, and thus much larger values of $P_{max}$ are required to reach a reliable estimate of $c_-$.
 
\section{ Incompressibility of the Gaffnian phase}\label{sec:inc}

Having established a close relationship of the ground state, the quasihole and quasielectron manifold between the Gaffnian phase and the Jain $\nu=2/5$ phase, we now attempt to synthesize these results and  turn our attention to the important issue of whether the Gaffnian state can be gapped in the thermodynamic limit, and if the topological nature of the Gaffnian phase is Abelian or non-Abelian.

 The numerical analysis of the Gaffnian model Hamiltonian $H_{\rm Gaff}$, as well as the connection to the non-unitary CFT, suggest that $ H_{\rm Gaff}$ is gapless in the thermodynamic limit~\cite{ReadViscosity,JolicoeurGaffnian}. The quasihole manifold of $H_{\rm Gaff}$, on the other hand, is confirmed to be non-Abelian in Ref.~\onlinecite{SimonGaffnian} (although such a calculation might be invalid in a gapless system).  However, as we argued in Sec.~\ref{sec:intro}, properties of a specific microscopic Hamiltonian are non-universal. A more relevant question we can ask is: does there exist a microscopic Hamiltonian such that (i) the ground state is topologically equivalent to the Gaffnian state; (ii) the ground state has a charge gap in the thermodynamic limit; (iii) the Gaffnian quasihole manifold is quasi-degenerate with the ground state in the thermodynamic limit.

For the LLL Coulomb interaction, it seems both (i) and (ii) are satisfied for the Jain $\nu=2/5$ state, but the Gaffnian quasihole degeneracy is lifted. We conjecture this is the only difference between the Gaffnian phase and the Jain $\nu=2/5$ phase, and the fundamental reason why the Jain $\nu=2/5$ phase could be Abelian. The question is if it is possible to tune the interaction so as to maintain the charge gap of the ground state, while at the same time making the Gaffnian quasihole manifold quasi-degenerate with the ground state. In this way, a non-Abelian topological phase at $\nu=2/5$ could be realised, which potentially would be adiabatically connected to the Abelian Jain state, in the sense that the ground state gap to charged or neutral excitations are maintained. We will now explore this issue from the perspective of the LEC construction.

It is clear, based on the specific LEC condition $\hat c_G$, that we can interpret the Gaffnian ground state, as well as all of the Gaffnian quasihole states, as made entirely of the $\left(5,2\right)\text{Type}$ Laughlin quasielectrons. In particular for the quasihole manifold, each of the quasihole state can contain a multiple of $\left(5,2\right)\text{Type}$ Laughlin quasielectrons, together with possibly the presence of Laughlin quasiholes. Thus the variational energies of these Gaffnian quasihole states are dominantly determined by the creation energies of $\left(5,2\right)\text{Type}$ Laughlin quasielectrons and Laughlin quasiholes, and the interaction between them. Both contributions depend on the microscopic Hamiltonian. With short range interactions (e.g., $V_1$ or Coulomb interaction in the LLL), the creation energy of $\left(5,2\right)\text{Type}$ Laughlin quasielectrons dominates. The interactions between them are relatively weak, and the creation energy of Laughlin quasiholes is negligible. We look at cases where the splitting of Gaffnian quasihole energies is mainly determined by the creation energy of $\left(5,2\right)\text{Type}$ Laughlin quasielectrons. If the creation energy of the quasielectrons (which we denote as $\Delta_{\left(5,2\right)}$) is large, the quasi-degeneracy of the Gaffnian quasihole manifold will be lifted. In contrast, with $\hat H_{\rm Gaff}$ the creation energy of $\left(5,2\right)\text{Type}$ Laughlin quasielectrons is zero.
\begin{figure}[htb]
\includegraphics[width=\linewidth]{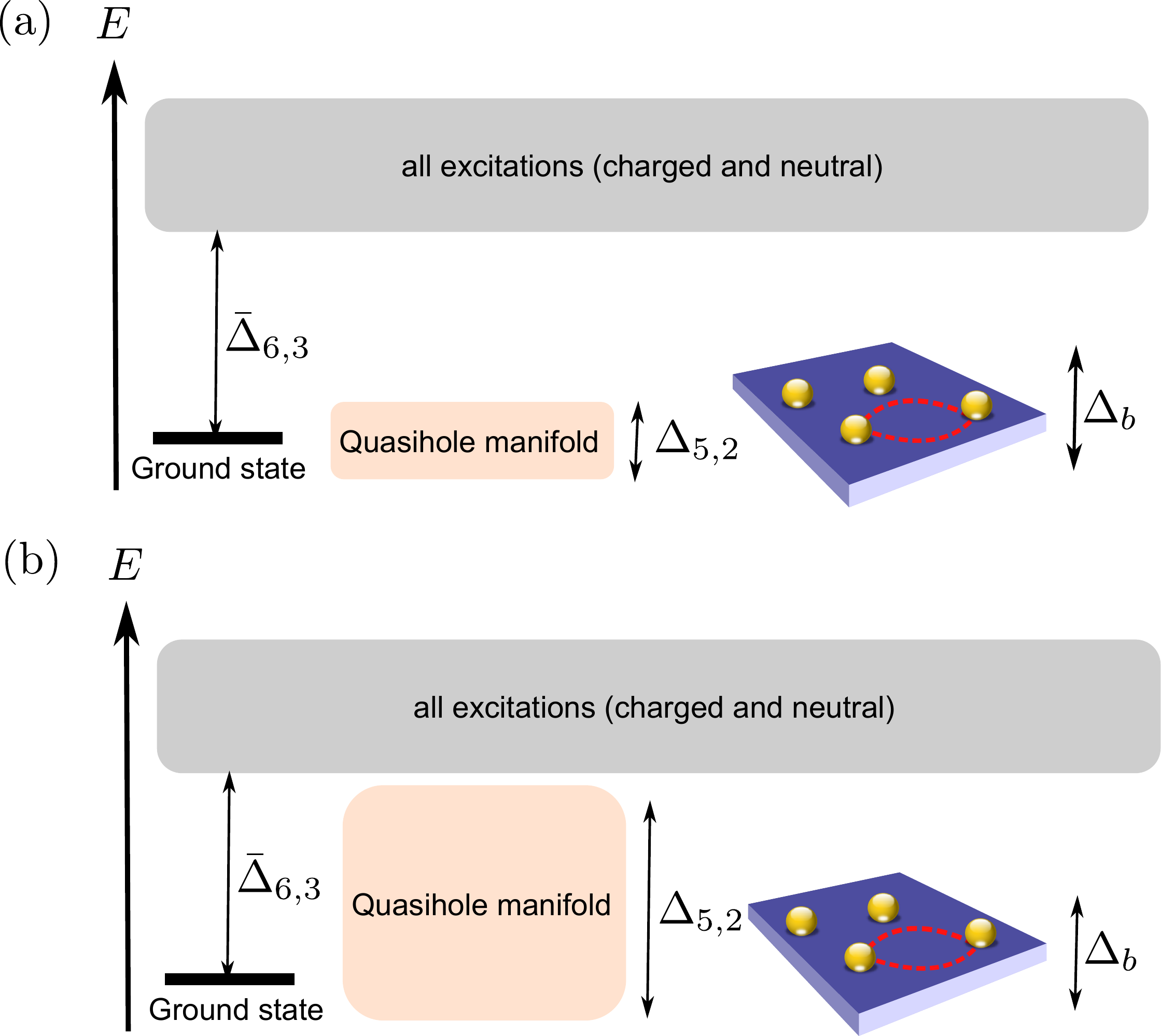}
\caption{Ground state charge gap, the energy splitting of the quasihole manifold, and the braiding energy scale. The incompressibility as well as the braiding outcome could be dominated by the self energies of different types of quasielectrons, depending non-universally on the microscopic Hamiltonians. (a) When $\bar{\Delta}_{6,3} \gg \Delta_b \gg \Delta_{5,2}$, we expect to see the non-Abelian statistics. (b) If the quasihole bandwidth is larger than $\Delta_b$, braiding experiment would show Abelian statistics. }
\label{fig2}
\end{figure} 

The incompressibility of the Gaffnian phase depends on the energy cost of the charged excitations on top of the Gaffnian ground state, see Fig.~\ref{fig2}. From the LEC construction, the quasielectron excitations (and thus the neutral excitations) can be classified as the $\left(6,3\right)\text{Type}$,  $\left(7,4\right)\text{Type}$ (i.e., defined by $\hat c=\{2,1,2\}\lor\{6,3,6\}, \{2,1,2\}\lor\{7,4,7\}$) etc., and the creation energies of these quasielectrons have to be non-zero in the thermodynamic limit, for the state to be incompressible. For realistic interactions in general, numerical simulations show the creation energy of the $\left(6,3\right)\text{Type}$ Gaffnian quasielectrons is the lowest, which we define as $\bar\Delta_{\left(6,3\right)}$. 

One should note $\left(5,2\right)\text{Type}$ Laughlin quasielectrons and $\left(6,3\right)\text{Type}$ Gaffnian quasielectrons are well-defined microscopic objects independent of the microscopic Hamiltonians. Their creation energies, on the other hand, depend on the microscopic Hamiltonians, which in turn controls the incompressibility of the Gaffnian phase, as well as if the phase is Abelian or non-Abelian. When adiabatically braiding  the quasiholes at $\nu=2/5$ in experiments, the action of braiding itself involves an energy scale of $\Delta_b$. Normally, adiabaticity implies $\Delta_b\rightarrow 0$, but in this limit, non-Abelian braiding is only possible if the quasihole manifold is exactly degenerate (especially the internal degeneracy when the quasihole locations are fixed). If we treat $\bar\Delta_{\left(6,3\right)}$ and $\Delta_{\left(5,2\right)}$ in Fig.~\ref{fig2} as the ground state gap and the quasihole broadening in general, then we only require adiabaticity to correspond to $\bar\Delta_{\left(6,3\right)}\gg\Delta_b$ in realistic implementations. When this is satisfied, the Gaffnian phase is only non-Abelian if $\Delta_b\gg\Delta_{\left(5,2\right)}$, otherwise the quasihole braiding will be Abelian, as assumed for the Jain $\nu=2/5$ phase.

\section{ Conclusions and outlook}\label{sec:conc}

We have presented a detailed comparison between the non-Abelian Gaffnian phase and the Abelian Jain phase at $\nu=2/5$, from the perspective of the LEC and CF constructions, respectively. The comparison involves not only the ground state, but also the low-lying elementary excitations such as quasiholes and quasielectrons. We have demonstrated that the Gaffnian ground state and the Jain $\nu= 2/5$ ground state are made of the same type of Laughlin quasielectrons. The low-lying quasielectron states of the Gaffnian phase also have high overlaps with and same counting as the Jain $\nu = 2/5$ quasielectron states. The Abelian Jain $\nu = 2/5$ quasiholes consist of only undressed CF quasielectrons of the Laughlin state in the CF picture. They again have the same counting and very high overlap with a subset of the non-Abelian Gaffnian quasihole states, which are made of undressed (5, 2) Type quasielectrons of the Laughlin state. In the CF picture, the dressed CF quasielectrons are excluded from the Jain $\nu = 2/5$ quasihole manifold when interactions are short range (e.g., the LLL-projected Coulomb interaction), which assigns higher variational energies to the dressed quasielectrons.

Based on our numerical analysis, which involves the counting of states and wave function overlaps, we have generalized the comparison between the Gaffnian and Jain $\nu=2/5$ phase to propose that topological properties of FQH phases should be defined without Hamiltonians or any local operators. Model Hamiltonians are useful tools but not fundamental for defining the topological aspects of FQH states. The important message here is that the topological properties of Gaffnian phase can be completely defined via the LEC conditions without invoking a specific Hamiltonian. In contrast, the Jain $\nu = 2/5$ phase does require a short range Hamiltonian, especially for the definition of the quasihole counting, which is crucial for the characterization of the phase to be Abelian. We thus propose that the two phases are ``topologically equivalent" at low energies, in the sense that \emph{all} topological indices computed by measurements on these two states are equal. For all the cases we have studied, the state counting from the LEC and the CF construction agrees, and the wave function overlaps decrease rather slowly with the system size. Even for states containing multiple quasielectrons, the wave function overlaps are very high despite quasielectron interactions and finite-size effects due to the intrinsic size of each quasielectron.

On the other hand, it is also important to study how robust these topological properties are in the presence of different types of Hamiltonians, which has experimental consequences. While it is common knowledge that the incompressibility (or the charged) gap is essential for the robust measurement of topological indices such as the filling factor or the topological shifts, the robustness of the quasihole degeneracy with realistic interactions has not been investigated in detail so far. The latter also seems to be independent of the incompressibility gap. If two Hamiltonians are adiabatically connected by the ground state incompressibility gap, it is still possible in principle for the quasihole degeneracy to differ significantly in the thermodynamic limit. In this paper we argue the Gaffnian description captures all of the topological properties of this FQH phase, including the quasihole manifold. In the LLL with the Coulomb interaction, which decays sufficiently fast,  the degeneracy of the quasihole manifold is not robust, and the CF description emerges as an effective theory, which captures the low-energy part of the Gaffnian phase, where excitations possess Abelian braiding properties.

Our work raises several interesting questions. For example, we have seen that by removing the assumption of short-range interactions, the Jain $\nu=2/5$ state supports a generalized quasihole manifold, $\bar{\mathcal{Q}}^J$, which is potentially even richer than the manifold of non-Abelian Gaffnian quasiholes. It would be interesting to understand the counting of states in this manifold and how they can be realized in a microscopic model. Further, it would be interesting to see if the relationship between the Gaffnian and the Jain $\nu=2/5$ state applies to the case between the LEC $\nu=3/7$ state and the Jain Abelian state at the same filling factor. For non-Abelian FQH phases, the presence of robust quasihole degeneracy is crucial for experimental realization. It is thus also worthwhile to examine the familiar Moore-Read state at $\nu=5/2$. With the three-body model Hamiltonian, the Moore-Read quasiholes are exactly degenerate and non-Abelian. In experiments, however, the realistic interaction is quite different from the model Hamiltonian. While many numerical calculations show that the ground state could be adiabatically connected to the model Hamiltonian, few studies have analyzed the robustness of the Moore-Read quasihole degeneracy in the presence of realistic Hamiltonians.

\begin{acknowledgments}
We thank Ajit C. Balram for useful discussions, Steve Simon for useful comments on the manuscript and clarifying the Gaffnian central charge, and especially Nicolas Regnault for generously sharing the entanglement spectrum of the Gaffnian state. This work is supported by the NTU grant for Nanyang Assistant Professorship.  ZP acknowledges support by EPSRC grant EP/R020612/1. Statement of compliance with EPSRC policy framework on research data: This publication is theoretical work that does not require supporting research data.

\end{acknowledgments}

\bibliography{fqhebib}
\end{document}